\documentclass[aps,prl,floatfix,amssymb,twocolumn,tightenlines,a4paper]{revtex4}
\usepackage{amsfonts}
\usepackage{amsmath}
\usepackage{color}
\usepackage{epsfig}
\usepackage{graphicx}
\usepackage{latexsym}
\usepackage{units}

\begin{document}
\title{Understanding photonic quantum-logic gates: \\ The road to fault tolerance}

\author{Till J. Weinhold$^{1-3}$, Alexei Gilchrist$^{4}$, Kevin J.~Resch$^{5}$, Andrew~C.~Doherty$^{1}$, Jeremy L. O'Brien$^{6}$, Geoffrey J. Pryde$^{3}$, and Andrew~G.~White$^{1,2}$}
\affiliation{$^{1}$Department of Physics and $^{2}$Centre for Quantum Computer Technology, University of Queensland, Brisbane QLD, 4072, Australia\\
$^{3}$ Centre for Quantum Dynamics, Griffith University, Brisbane 4111, Australia\\
$^{4}$ Centre for Quantum Computer Technology and Department of Physics, Macquarie University, Sydney, NSW 2109, Australia\\
$^{5}$Institute for Quantum Computing and Department of Physics \& Astronomy, University of Waterloo, Waterloo, ON,  N2L 3G1 Canada\\
$^{6}$H. H. Wills Physics Laboratory and Department of Electrical and Electronic Engineering, University of Bristol, Bristol BS8 1UB, UK.}

\begin{abstract}
\noindent Fault-tolerant quantum computing requires gates which function correctly despite the presence of errors, and are scalable if the error probability-per-gate is below a threshold value. To date, no method has been described for calculating this probability from measurements on a gate. Here we introduce a technique enabling \emph{quantitative} benchmarking of quantum-logic gates against fault-tolerance thresholds for \emph{any} architecture. We demonstrate our technique experimentally using a photonic entangling-gate. The relationship between experimental errors and their quantum logic effect is non-trivial: revealing this relationship requires a comprehensive theoretical model of the quantum-logic gate. We show the first such model for any architecture, and find multi-photon emission---a small effect previously regarded as secondary to mode-mismatch---to be the dominant source of logic error. We show that reducing this will move photonic quantum computing to within striking distance of fault-tolerance.
\end{abstract}
\maketitle

The increase in computational power of classical computers is driven by the miniaturisation of their components, which is inexorably approaching the quantum level. While heat and noise ultimately limit the computational power of classical computers, embracing the features of a quantum mechanics opens the door to a novel computing paradigm: quantum computation. Here two-level quantum systems are used as carriers of the information and are thus called quantum bits or qubits for short. Their unique features, including being encoded in superposition states and forming entanglement with other qubits, leads to capabilities impossible with traditional computation \cite{MikeIke}, ranging from efficient modelling of quantum systems \cite{CTDThesis}, to defeating widely-used encryption protocols \cite{Shor94}. Multiple architectures are being investigated for their potential to realise quantum computation \cite{ion1,ion2,photon1,photon2,photon3,superconductor,atom}, one of the most promising being photonic quantum computing utilising measurement induced non-linearities \cite{KLM} to realise the essential entangling gates.

While significant progress has been made and proof-of-principal implementation of some of the intriguing algorithms \cite{Grover,ShorExp1,ShorExp2} have been demonstrated, all of the demonstrated gates suffered a variety of errors due to intrinsic noise. To scale up and realize a quantum computer large enough to harvest the full benefits of this paradigm, one needs fault-tolerant gates \cite{Shor95,Steane}, that is gates that work correctly despite the presence of intrinsic errors. A gate is said to be scalable if the error probability-per-gate lies below a threshold value. A further difficulty is that fault-tolerant benchmarking identifies the magnitude, but not the source of errors. That is, obtaining the error probability-per gate gives information about how close---or far!---a gate is from being fault-tolerant, but no indication on how to get there. To reveal the sources of the noise, and their impact on the error-probability-per-gate, one needs to develop a comprehensive model of the logic gate. As each quantum computing architecture suffers different noise sources these models are architecture specific. 

Here we present the first method that allows benchmarking of experimental gates against fault-tolerant thresholds, and demonstrate it with photonic quantum-logic gates. Further, to identify the sources of the noise and their impact on the error-probability-per-gate, we develop a comprehensive model of the quantum-logic gate which uses only parameters derived from our experiment. We find that emission of multiple photons in the same spatio-temporal mode is the main source of error in photonic quantum-logic gates: alleviating this effect will place photonic quantum computing on the doorstep of one of the recently derived fault-tolerance thresholds \cite{Knill05}.

\begin{figure*}[!htb]
\begin{center}
\includegraphics[width=\linewidth]{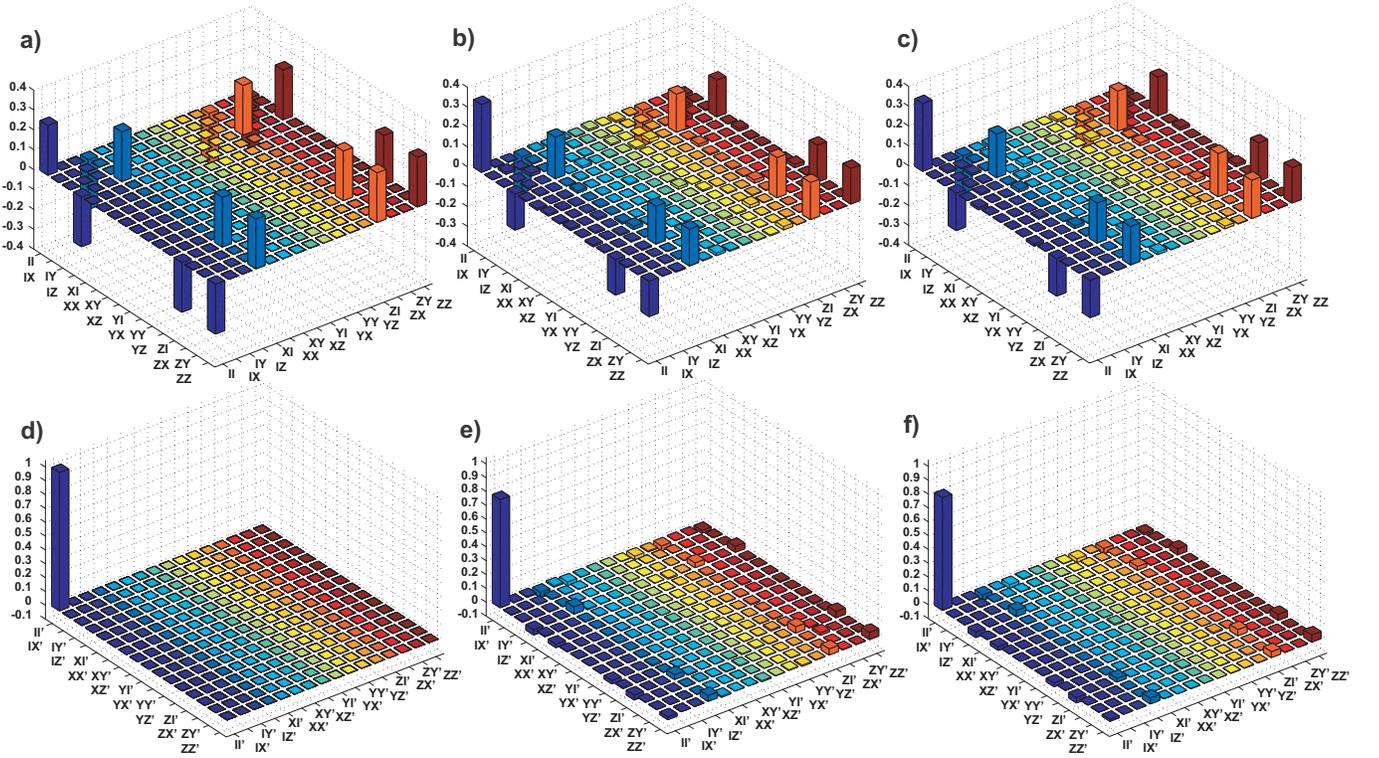}
\vspace{-0.6 cm}
\caption{$\chi$-matrices for the bit-flipped controlled-\textsc{z} gate, $({-}$\textsc{i$_{1}$i$_{2}$}+\textsc{i$_{1}$z$_{2}$}+\textsc{z$_{1}$i$_{2}$}+\textsc{z$_{1}$z$_{2}$}), where the subscript indicates which qubit the operator acts on. a)-c) are traditional Pauli-basis, d)-f) are gate-basis, e.g. $(\textsc{i}_{1}\textsc{i}_{2})'{=}\textsc{i}_{1}\textsc{i}_{2}{\otimes}\textsc{cz}_{12}$; the first element represents ideal gate operation, all other elements indicate errors. In d)-f), the population of the first diagonal element is the process fidelity $F_{p}$; the gate error is at least as large as the combined population of the remaining diagonal elements, $1{-}F_{p}$. a), d) \emph{Ideal gate}. b), e) \emph{Experimentally-measured gate}. Process and average-gate fidelities with the ideal, $\{F_{p},\overline{F}\}$, are respectively $\{78.2,82.5\}{\pm}1.5$\%. c), f) \emph{Modelled gate}. Fidelities with the ideal are $\{81.4,85.1\}$; with experiment $\{96.7,97.4\}{\pm}1.5$\%. \emph{n.b.} Only real parts are shown; imaginary parts are ideally zero, for both experiment and model they are ${<}4$\%,  with an average value of ${<}0.3\%$.}
\label{chis}
\end{center}
\vspace{-0.8 cm}
\end{figure*}

\vspace{-4mm}
\section{Benchmarking against fault-tolerance thresholds}
\noindent Calculating fault-tolerance thresholds is an exceedingly difficult theoretical problem, with the result depending on essential assumptions regarding the types of errors \cite{Gott07}. Typically, fault-tolerant models consider a gremlin \cite{Dahl} which can introduce errors on any qubit, at any stage in the circuit, with probability $\varepsilon$. Two common approaches are to restrict the errors and derive the threshold numerically, or alternatively to allow a more generalised error model and find the threshold analytically---the former generally yields higher thresholds than the latter. The model of Knill \cite{Knill05} is an example of the former: the errors are uncorrelated---random, independent Pauli operations---and error thresholds as high as $\varepsilon_{0}{=}$3-6\% per gate are shown to be tolerable (albeit with large encoding overheads). An example of the latter approach is the model of Aliferis, Gottesman, and Preskill \cite{agp}, which allows for adversarial independent stochastic noise, where the errors are \emph{any} physical process that can be described by a completely-positive map---which includes entangled errors!---deriving a threshold of $\varepsilon_{0}{=}2.73 \times 10^{-5}$. Experimentally, such correlated errors are an unlikely scenario.

Regardless of the method used to derive the threshold, bridging the gap between a predicted threshold and a given experimental gate implementation requires quantitatively determining the implemented quantum process \cite{qpt}. Quantum processes can be compactly represented by the $\chi$-matrix, a table of process-probabilities and the coherences between them (analogous to the density matrix, $\rho$, which is a table of state measurement outcomes and the coherences between them). The overlap of the ideal with an experimentally determined $\chi$-matrix is the \emph{process fidelity}. Process fidelity is the basis of several performance metrics \cite{gln}, however none of these metrics provide an error probability-per-gate to allow direct comparison with fault-tolerance thresholds. Here we show how to obtain the gate error from the $\chi$-matrix.

The gremlin has a probability, $\varepsilon$ of replacing the correct process, $\chi_{\mathrm{ideal}}$, with an error process, $\chi_{\mathrm{gr}}$,
\begin{equation}
(1{-}\varepsilon) \chi_{\mathrm{ideal}} + \varepsilon \chi_{\mathrm{gr}},
\end{equation}
where for example, $\chi_{\mathrm{gr}}$ is restricted to Pauli processes in the Knill mode, or is any completely-positive process in the Aliferis, Gottesman, and Preskill model. We assume the gremlin is adversarial and attribute to it all observed incoherent errors in the experiment. We want to find the minimum-$\varepsilon$, $\varepsilon^*$, such that  $\chi_{\mathrm{gr}}$ still represents a physical, trace-preserving process---that is, we require that all eigenvalues are non-negative and that $\mathrm{Tr}_A\chi_{\mathrm{gr}}{=}\textsc{i}/d$. where \textsc{i} is the identity operator, and $d$ is the dimension of the $\chi$-matrix. The optimisation we want to solve is the following,
\begin{equation}
\min \; \varepsilon, \mbox{ such that } \varepsilon\chi_{\mathrm{gr}} {=}  \chi_{\mathrm{exp}} {-} (1{-}\varepsilon)\chi_{\mathrm{ideal}} {\ge} 0.
\label{primal}
\end{equation}
Note that the partial-trace condition for $\chi$ matrices is automatically satisfied since both $\chi_{\mathrm{exp}}$ and $\chi_{\mathrm{ideal}}$ represent physical process matrices. This optimisation is in the form of a \emph{semidefinite program}, a convex optimisation problem which enjoys several advantages, such as being particularly amenable to numerical solution, and that every local optimum is equal to a global optimum \cite{VaBbook}. 

Commonly, $\chi$ is represented in the Pauli-basis, tensor-products of $\{$\textsc{i},\textsc{x},\textsc{y},\textsc{z}$\}$, where \textsc{x},\textsc{y},\textsc{z} are the Pauli-spin operators, e.g. the controlled-\textsc{z} (\textsc{cz}) gate in Fig.~\ref{chis}a)-c).  A useful alternative representation is the \emph{gate} basis, where the ideal gate operation multiplies each Pauli basis operator, Fig \ref{chis}d)-f). The first element now represents ideal gate operation, e.g. \textsc{cz}$_{12}$.\textsc{i}$_{1}$\textsc{i}$_{2}$: the population \emph{is} the process fidelity between the ideal and measured matrices \cite{qpt}, $F_{p}$. The other diagonal elements represent combinations of Pauli errors after an ideal gate operation. If the first element in the gate basis shares no coherences with any other element then the sum of the remaining populations gives the gate error, i.e. $\varepsilon^{*}{=}1{-}F_p$ (see supplementary material). This is the case for gates with errors described by the Knill error model. 

More generally, the errors may share coherence with the ideal process, and so $1{-}F_p$ is at best a lower bound for the minimum error-probability per gate, $\varepsilon^{*}$. In principle it would be possible to easily solve the optimisation (\ref{primal}) numerically. However, in current state-of-the-art process tomography, the experimental matrices are reconstructed via a maximum-likelihood technique that often results in one or more zero eigenvalues, leading to the unhelpful solution $\varepsilon^{*}{=}1$. To circumvent this we use a dual optimisation technique, and add a realistic noise process of variable strength, $\delta$. As we are adding noise this provides us with an \emph{upper} bound to $\varepsilon^{*}$ (see supplementary material). Thus, for any measured gate, our procedure provides bounds for the minimum error-probability per gate, allowing direct comparison to the fault-tolerant thresholds, $\varepsilon_{0}$.

\begin{figure}[!t]
\begin{center}
\includegraphics[width=\linewidth]{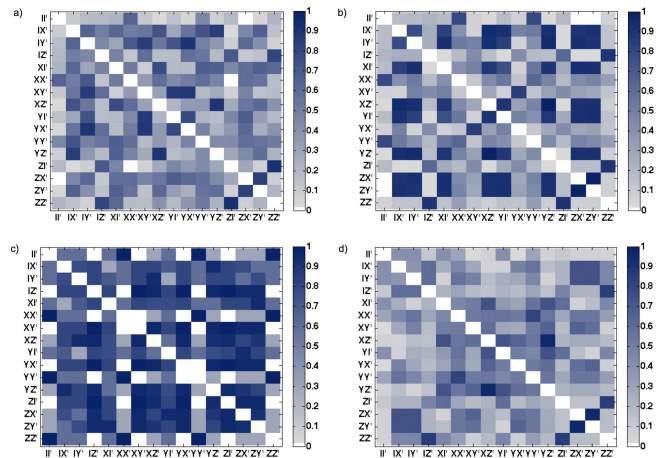}
\vspace{-2 mm}
\caption{Coherence matrices showing the degree of coherence in the gate basis, $\mathcal{C}$, for: a) the experiment; b) the full model; c) modelling only the effect of imperfect circuitry; and d) modelling only the effect of higher order photon numbers. $\mathcal{C}$ varies between 0, no coherence, and 1, full coherence. The ideal coherence matrix is zero, as for a $\chi$-matrix in the gate basis all population is in the first element and there are no off-diagonal elements. Note the striking similarity between the experiment and model coherence matrices. The model predicts blocks of coherence not observed experimentally, varying the individual error sources let us investigate their cause. c) shows that the majority of the predicted strong coherences are due to non-ideal circuitry; d) shows that that the strong coherences observed in the experiment, e.g. \textsc{iz'-zz'}, \textsc{xz'-yz'}, \textsc{zi'-zz'}, and \textsc{zx'-zy'}, are primarily due to the higher-order terms. Since the experiment suffers decoherence due to mode mismatch we expect it to have lower coherences than the model, as is observed. \vspace{-12mm}
}
\label{doc}
\end{center}
\end{figure}

We illustrate these techniques by applying them to a photonic controlled-\textsc{z} gate. There are known paths for achieving scalable, fault-tolerant, quantum computation using these gates, either via the circuit model \cite{KLM} or the measurement-based model  \cite{Nielsen-Cluster}. In practice this will require many \emph{independent} photon sources: the majority of previous demonstrations of entangling photonic gates have used \emph{dependent} photons generated in the same down-conversion event. These are not a suitable model for future photon sources as they share correlations due to their paired birthing. Here we instead use independent photons generated in separate down-conversion events, which  share no such correlations, to implement the controlled-\textsc{z} gate previously described in Refs \cite{UQCZ,MunichCZ, JapanCZ} (see supplementary material). We detect one photon from each down-conversion event to herald its sister photon which is input to our gate.  The $\chi$-matrix is measured using an automated quantum process tomography system \cite{qpt,UQCZ}; we follow the established practice of searching over all single qubit unitaries to rotate the experiment as close as possible to the ideal \cite{UQCZ}, i.e. we maximise the average gate fidelity, $\overline{F}$. The real part is shown in Figs \ref{chis}b) and e) in the Pauli and gate bases, respectively. 

\begin{figure*}[!t]
\begin{center}
\includegraphics[width=1.0 \linewidth]{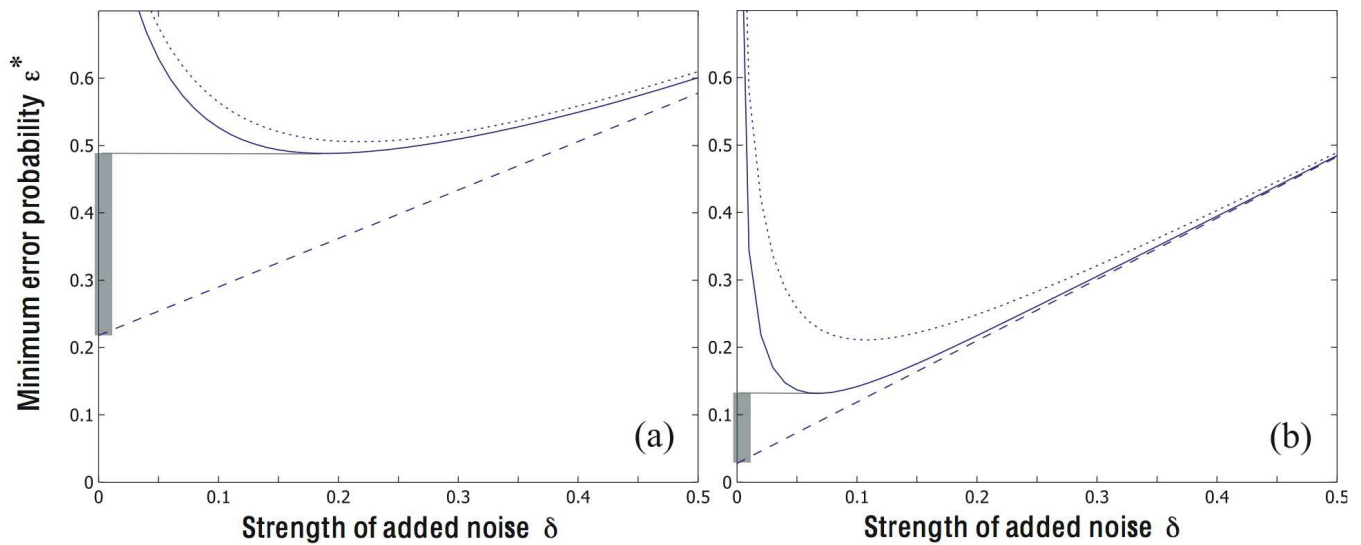}
\vspace{-3 mm}
\caption{Grey bar indicates range of minimum error-probability per gate, $\varepsilon^{*}$. The lower bound (straight dashed line) is $1{-}F_{p}$.  The upper bound is derived through the deliberate addition of noise with varying strength $\delta$, so that the reconstructed experimental process, is of the form $(1{-}\delta)\chi_{\mathrm{exp}}{+}\delta\chi_{\mathrm{noise}}$ (see supplementary material). The dashed curve results from adding depolarising noise; the solid curve from optimising the form of the noise. These curves are upper bounds on the true $\varepsilon^{*}$ since we are deliberately adding extra noise to the experimental case. a) For the experimentally measured gate shown in Fig.~\ref{chis}b), e), the error per gate probability is bound by $21.8\%{\leqslant}\varepsilon^{*}{\leqslant}48.8\% $. b) These bounds reduce drastically to $2.8\%{\leqslant}\varepsilon^{*}{\leqslant}13.2\% $ when modelling the removal of higher-order terms, but keeping the measured beamsplitter values. Note that the bounds on the gate error tighten considerably.\vspace{-9mm}
}
\label{bounds}
\end{center}
\end{figure*}

The gate has high fidelity with the ideal, the population of the first element in Fig.~\ref{chis}e) is $F_{p}{=}78.2{\pm}1.5$\%. Notably, this fidelity is many standard deviations less than those achieved with dependent photon inputs \cite{qpt,UQCZ}, $F_{p}{=}87$\%, indicating the presence of an extra noise source. As can be seen from Fig.~\ref{chis}e), the distribution of errors is not random, nor are the coherences zero. The coherences can be quantified using the \emph{degree of coherence}, $\mathcal{C}_{ij}{=}|\chi_{ij}|(1{-}\delta_{ij})/\sqrt{\chi_{ii}\chi_{jj}}$, where $i,j$ are indices of the $\chi$-matrix. Entries in the resulting coherence matrix vary between 0, no coherence, and 1, maximal coherence. The coherence matrix for the experiment is shown in Fig.~\ref{doc}a), clearly showing non-zero coherences---the Knill error model is not appropriate here! The lower bound to the minimum gate error is thus $\varepsilon^{*} {\geqslant} 1{-}F_{p} {=} 21.8{\pm}1.5$\%; the upper bound provided by the convex optimisation procedure, Fig.~\ref{bounds}a, is $\varepsilon^{*} {\leqslant} 48.8$\%. The upper bound is quite loose due to our procedure, which adds extra noise in order to function. Future theoretical advances that do not require this will tighten the bounds substantially. Nevertheless, we are clearly far from any fault-tolerant threshold.

\section{Error model for a photonic quantum-logic gate}
\noindent To achieve fault-tolerant computing the gate errors must be reduced. The $\chi$-matrix alone does not provide enough information to do this, since it tells us the probability of error, but does not identify the noise sources or their strengths. Noise sources vary greatly between different physical systems, and an important first-step is to measure an experimental error budget \cite{ion1}. However, in general there is a non-trivial relationship between experimental errors and their effect in quantum logic---a comprehensive \emph{architecture-dependent} model of the quantum gate is required, incorporating a realistic physical description of possible noise sources. In photonic quantum computing the major source of noise has previously been identified as mode mismatch \cite{photon1,qpt,UQCZ}, albeit on gates using dependent photons. We use independent photons here because as highlighted above, dependent photon sources are not scalable. Experimentally, we built a fibre-coupled gate where we could vary the input between dependent and independent photon sources with \emph{no} change in spatial or temporal mode mismatch.

\begin{figure}[!t]
\begin{center}
\includegraphics[width=1 \columnwidth]{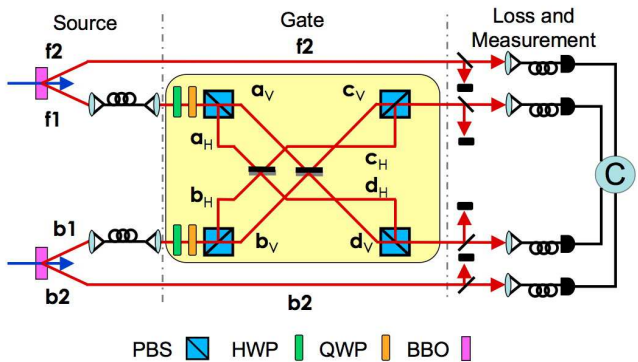}
\vspace{-0.6 cm}
\caption{Architectural model of noise in a linear photonic gate. Three separate physical sources of noise are considered: the downconversion photon source, including intrinsic higher-order photon terms; gate imperfections, where gate reflectivities vary from the ideal; and photon loss, modelled as a beamsplitter preceding the measurement. The photons from the source are fibre-coupled, polarised and sent through quarter- and half- wave plates to the gate. Ideally the reflectivities in the gate are $\eta_{H}{=}1/3$, $\eta_{V}{=}1$; in practice these are realised on a single partially-polarising beamsplitter with collinear inputs. The gate outputs are sent to quarter- and half- wave plates, and polariser (not shown) for tomography. The photons are detected by avalanche photodiodes, fourfold coincidences signal gate operation. 
}
\label{Model}
\end{center}
\vspace{-0.8 cm}
\end{figure}

To describe this, we construct a model that captures the difference between dependent and independent photon sources in our experiment: the photon source characteristics, including higher-order photon terms; imperfections in the optical architecture; and photon loss, Fig. \ref{Model}. (In effect, our model accounts for all known error sources except mode mismatch). The source is described by a Taylor-expansion of the interaction Hamiltonian for parametric down-conversion (see supplementary material). We consider only terms generating one pair per downconverter, or $2{+}1$ or $1{+}2$ pairs. As at least one photon pair is required in either source to trigger a fourfold event, terms generating only one pair, or all pairs in one mode can be ignored, as can terms describing the generation of four pairs due to their exceedingly low probability. Conceptually, two of the spatial modes are split into polarisation modes which experience different non-classical interferences; in practice these are realised collinearly on a partially-polarising beamsplitter. Without loss of generality, all photon losses can be modelled at a single stage as the circuit is linear: we choose to model loss just before the projective measurements. In an otherwise ideal gate, loss does not contribute, as a valid four-photon event would not be registered. When non-single photons are injected, a valid signal can occur even if one or two photons were lost. We find that due to altered interference behaviour, high loss tends to emphasise the negative effects of non-ideal photon sources. The resulting model has 8 parameters, each of which is determined from experiment (see supplementary material for the detailed derivation).

Model $\chi$-matrices are obtained by using the same set of measurements and process tomography procedure used in the experiment. As Figs 1c), f) show, our model is highly accurate, with a process fidelity between the model and the experiment of $F_{p}{=}96.7{\pm}1.5$\%. We attribute the difference of $3.3$\% as due to mode mismatch, the only known error source not included in our model. Mode mismatch may be spatial, spectral, or temporal: in all cases it has the same effect of increasing distinguishability of the photons. In our current model there is still some cancellation of higher-order photon terms due to nonclassical interference, mode mismatch will degrade these, reducing the model values even closer to the experiment. The agreement between model and experiment goes well beyond high fidelities, with good agreement between the measured and modelled coherence matrices, Figs~\ref{doc}a), b).

\begin{table}[!t]
\begin{tabular}{|l|c|c|c|} \hline
Model Settings    & $\{F_{p},\overline{F} \}_{ideal}$ [\%] & $\{ F_{p},\overline{F} \}_{exp}$ [\%] & $1{-}F_{p}$ [\%]\\ \hline \hline 
ideal 			&$100,100$ &$78.2,82.5$	& 0		\\
loss				&$100,100$ &$78.2,82.5$	& 0		 \\
gate				&$97.2,97.8$ &$80.2,84.2$	& $2.8$	 \\
gate+loss			&$97.2,97.8$ &$80.2,84.2$	& $2.8$	 \\
source			&$93.2,94.6$ &$92.1,93.7$	& $6.8$	  \\
source+gate		&$88.0,90.4$ &$93.6,94.9$	& $12.0$ \\
source+loss		&$87.2,89.8$ &$94.4,95.5$	& $12.8$ \\
source+gate+loss	&$81.4,85.1$ &$96.7,97.4$	& $18.6$ \\
\hline \hline
experiment		&$78.2,82.5 \ {\pm} 1.5$ & $100,100$ 	& $21.8{\pm} 1.5$ \\
\hline
\end{tabular}
\caption{Process and average-gate fidelities with the ideal, $\{F_{p},\overline{F}\}$, and absolute lower bounds for the minimum error probability-per-gate, $\varepsilon^{*}{\geqslant}1{-}F_{p}^{ideal}$ for model and experiment. Terms that can be turned on in the model are: \emph{loss}, given by measured losses in the experiment; \emph{gate}, given by measured beamsplitter reflectivities in the experimental gate; and \emph{source}, higher-order photon terms based on measured rates and loss. In the \emph{ideal} case there is no loss or  higher-order photon terms, and ideal beamsplitter reflectivities. \vspace{-6mm}
}
\label{error budget}
\end{table}

The striking advantage of architectural models is that by varying the parameters between ideal and measured values we can explore the contributions of each error source. Table \ref{error budget} summarises the results for our photonic model. There are several things to note here. Firstly, it is clear that loss has no effect in the absence of higher-order photon terms. This is expected as ideally the gates have two single-photon inputs and either non-destructive \cite{DowlingQND} or destructive coincident detection: losing a photon does not give a valid signal and thus leads to no errors. Secondly, the combined effect of imperfect beamsplitter reflectivities and loss alone are rather small, $\Delta \{F_{p}, \overline{F} \}{=}\{2.8,2.2\}$\%; these errors are highly coherent as shown in Fig.~\ref{doc}c). Deriving the fault-tolerance related error bounds for this result from our error model predicts that $2.8 {\leqslant} \varepsilon^{*} {\leqslant} 13.2\%$, as shown in Fig.~\ref{bounds}b). Thirdly, adding in the higher-order photon terms leads to a large degradation, $\Delta \{F_{p}, \overline{F} \}{=}\{15.8,12.7\}$\%. If we turn on only the higher-order photon terms, we see that the large introduced errors are mostly incoherent, Fig.~\ref{doc}d), the combined effect of source and gate imperfections on the coherences is shown in Fig.~\ref{doc}b). Finally, the difference in fidelities with the ideal between the full model and experiment are small, $\Delta \{F_{p}, \overline{F} \}{=}\{3.2,2.6\}{\pm}1.5$\%. The model is highly accurate: we note again that there are no free fit parameters, all values being determined directly from experiment.

These results show that higher-order photon terms are a major source of noise in photonic entangling-gates. At first glance, this is surprising since we use heralded single photons given that multiple pairs are emitted in only $1.2{-}3.2\%$ of downconversion events. Although unlikely, these events are significant because they can cause errors in multiple ways. Consider 2+1 photons entering the gate: 1) if the single photon is lost, a coincident detection can still occur despite the lack of necessary nonclassical interference; 2) if no photons are lost, the ensuing nonclassical interference has the wrong visibility  \cite{Sanaka} ; and 3) when an output is populated by more than one photon, this alters the individual detection efficiencies, leading to apparently higher populations of some output states.

There is a world-wide effort into developing true single photon sources \cite{spsexps} and photon-number-resolving detectors \cite{NRdets}; and there is demonstrated capability at making higher precision optics than used here. As we reduce the parameters for these terms in our model, it rapidly converges to the ideal, leaving mode mismatch as the dominant source of gate error. Mode mismatch errors are incoherent \cite{Rohde}, meaning that there will be no coherences in the gate basis and the minimum error-probability per gate is simply given by $\varepsilon^{*}{=}1{-}F_{p}$ as discussed above.  From the difference between the measured gate, and model, Table \ref{error budget}, we estimate $\varepsilon^{*}{=}3.2{\pm}1.5$\%---moving photonic quantum computing squarely within the range of the fault-tolerant error threshold of $3{-}6\%$  predicted by Knill. 

Our findings highlight the significant role of higher-order photon terms in photonic quantum information. Previous studies had considered the effects in quantum communication \cite{QComm1}, concluding that heralded downconversion would be a promising approach \cite{QComm2}. As we show here, even with heralded photon sources, higher-order photon terms are the leading error source in quantum logic. This result was only found by using a comprehensive, architecturally-dependent, error model that teased out the nonlinear relationship between experimental errors and quantum-logic errors. This is a general procedure, and will benefit any quantum information architecture in identifying the source and importance of experimental errors for quantum logic.

This work was supported in part by the Australian Research Council Discovery, Centre of Excellence, and Federation Fellow programs, and the IARPA-funded U.S. Army Research Office Contract No. W911NF-05-0397.

\end{document}


\title{Supplementary material. Understanding photonic quantum-logic gates: \\ The road to fault tolerance}

\author{Till J. Weinhold$^{1-3}$, Alexei Gilchrist$^{4}$, Kevin J.~Resch$^{5}$, Andrew C. Doherty$^{1}$, Jeremy L. O'Brien$^{6}$, Geoffrey J. Pryde$^{3}$, and Andrew G. White$^{1,2}$}
\affiliation{$^{1}$Department of Physics and $^{2}$Centre for Quantum Computer Technology, University of Queensland, Brisbane QLD, 4072, Australia\\
$^{3}$ Centre for Quantum Dynamics, Griffith University, Brisbane 4111, Australia\\
$^{4}$ Centre for Quantum Computer Technology and Department of Physics, Macquarie University, Sydney, NSW 2109, Australia\\
$^{5}$Institute for Quantum Computing and Department of Physics \& Astronomy, University of Waterloo, Waterloo, ON,  N2L 3G1 Canada\\
$^{6}$H. H. Wills Physics Laboratory and Department of Electrical and Electronic Engineering, University of Bristol, Bristol BS8 1UB, UK. \vspace{-6mm}}
\maketitle

\noindent \textbf{Chi matrices}. The $\chi$-matrix acts as the transfer matrix for \emph{any} input state $\rho_{in}$, where $\rho_{out}{=}\sum_{i,j}\chi_{i,j} \hat{A}_{i} \rho_{in} \hat{A}^{\dagger}_{j}$, and $\hat{A}$ are the basis operators acting on $\rho_{in}$.  The \emph{elementary basis} of operators is a convenient mathematical representation in which the $\chi$-matrix is obtained via the Jamiolkowski isomorphism $\chi {=} I\otimes \mathcal{E}(|\psi\rangle\langle\psi|)$, where $|\psi\rangle{=}\sum_j|jj\rangle/d$ is a maximally-entangled state between the basis vectors spanning the space $\mathcal{H}_A$ where the process $\mathcal{E}$ operates, and a copy of that space $\mathcal{H}_B$. In this basis a physical process (completely-positive trace-preserving)  is a positive-semidefinite matrix with partial trace $\mathrm{Tr}_A \chi{=}I/d$. For processes $\chi$ and $\sigma$, the process fidelity is $F_{p}(\sigma,\chi)\equiv Tr^{2}(\sqrt{\sqrt{\sigma} \chi \sqrt{\sigma}})$.

\vspace{1mm}
\noindent \textbf{Calculating gate errors}. The \emph{primal} optimisation,
\begin{equation}
\min \; \varepsilon, \mbox{ such that } \varepsilon\chi_{\mathrm{gr}} {=}  \chi_{\mathrm{exp}} {-} (1{-}\varepsilon)\chi_{\mathrm{ideal}} {\ge} 0,
\label{primal}
\end{equation}
possesses a \emph{dual} optimisation problem, 
\begin{align}
\max \;& d{=}\mathrm{Tr}\chi_{\mathrm{ideal}}Z{-}\mathrm{Tr}\chi_{\mathrm{exp}}Z,\nonumber \\
\mathrm{such\;that\;} & Z{\ge}0, \; \mathrm{Tr}\chi_{\mathrm{ideal}}Z{=}1,
\label{dual}
\end{align}
where the optimisation variable is a hermitian matrix $Z$. The primal and dual problems are related by a condition known as weak-duality, which asserts that any solution $\varepsilon$ of the primal problem is always greater than a solution $d$ of the dual problem, and in particular these sandwich the optimal solutions $\varepsilon^*$ and $d^*$: $\varepsilon {\ge} \varepsilon^{*} {\ge} d^* {\ge} d$. 

The dual problem can often be used to derive lower bounds on the primal optimum. For instance, the trial solution $Z{=}\chi_{\mathrm{ideal}}$ is a valid solution, and hence $\varepsilon^{*}{\ge}d{=}1{-}F_p$. In general this solution is not optimal so the bound is not saturated, this can be seen by checking the conditions for the primal problem when $\varepsilon{=}1{-}F_p$. The bound \emph{is} saturated if, in the gate basis, the first element of $\chi_{\mathrm{exp}}$, corresponding to $F_p$, shares no coherences with any other element---that is, $\chi_{\mathrm{exp}}$ has a block diagonal structure with two blocks, a $1{\times}1$ block followed by the remainder. The Knill error model is an example.  In such cases, the optimal solution is $\varepsilon^{*}{=}1{-}F_p$.

In principle it would be possible to easily solve the optimisation (\ref{primal}) numerically. However, in current state-of-the-art process tomography, the experimental matrices are reconstructed via a maximum-likelihood technique that often results in one or more zero eigenvalues. (An alternative technique based on Bayesian analysis, not yet in widespread use, may avoid this\cite{Blume}). Since in general it is not guaranteed that the expectation for the corresponding  eigenvectors with $\chi_{ideal}$ are also zero, and remembering that the gremlin process is required to be complete positive, $\varepsilon^{*}{=}1$ may be the only valid solution.  Clearly this is not a useful upper bound for gate errors! 

To overcome the trivial solutions imposed by the zero eigenvalues in the reconstructed $\chi$ matrix, we deliberately add noise to the process. By choosing the noise so that it could be generated and added with high fidelity in the lab, the resulting $\varepsilon^{*}$ will be an upper bound as we could add this noise in practice. We will consider noise generated by the following procedure --- with probability $\delta$ we replace the gate output with a fixed state, if this is the maximally mixed state, it adds depolarising noise.  In general we will optimise over the fixed state for this noise process $\chi_{noise}$ also, the total optimisation still being a semi-definite program.  Fig.~2 plots the upper and lower bounds for the measured and modelled gates. Our procedure allows the direct estimation of gate errors for a given experimental gate.

\vspace{1 mm}
\noindent \textbf{Experiment}. Our photon source uses a femtosecond-pulsed Ti:Saph laser at $820$nm, upconverted in a bismuth borate crystal (BiBO) to bi-directionally pump degenerate type-I parametric downconversion in a beta-barium-borate crystal (BBO). The pump is focussed so that it has different waist sizes for forward and backward downcoversion---mimicking the effect of independent photon sources with differing brightnesses. All four downconversion beams are fibre coupled; one photon of each pair (modes \textbf{f2},\textbf{b2}) is immediately detected  to herald its partner (\textbf{f1},\textbf{b1}), which is injected into the controlled-\textsc{z} gate. The gate is described in detail in Ref.~22, and is shown schematically in Fig.~4, where the horizontal and vertical interactions are realised with a single custom-made partially-polarising beamsplitter (PPBS). The asymmetric action and thus losses of the gate need to be balanced. In Ref.~22 additional PPBSs were used, here we pre-bias the input states. While the former requires no knowledge of the input state, the pre-biasing is input state dependent but yields higher count rates.  Horizontally-polarised modes experience three times more loss than vertical one. For pure states in this basis, this leads merely to different integration times to obtain the same counting statistics. For superposition state however, we need to inject an appropriately pre-biased state, i.e. $|D^{\prime}\rangle{=}(\sqrt{3}|H\rangle{+}|V\rangle)/2$ becomes the balanced diagonal state $|D\rangle{=}(|H\rangle{+}|V\rangle)/\sqrt{2}$ after reflection of a PPBS. The polariser, half- and quarter- waveplates before (after) the PPBS allow the controlled input (detection) of any desired state. All four detectors are fibre-coupled silicon avalanche photodiode detectors (Perkin-Elmer AQR-14-FC); fourfold coincidences are obtained by converting the TTL outputs to NIM via a constant fraction discriminator (ORTEC 935), and then inputting the NIM pulses to a quad-logic card (ORTEC CO4020).

\vspace{1 mm}
\noindent \textbf{Model}. To describe the photon creation we take the interaction component of the Hamiltonian for down-conversion as,
\begin{equation}
\mathbf{H} = A_{f} \mathbf{f1}^{\dagger} \mathbf{f2}^{\dagger} \mathbf{p}_{f} + A_{b} \mathbf{b1}^{\dagger} \mathbf{b2}^{\dagger} \mathbf{p}_{b} + h.c.,
\end{equation}
where $A_{b}$, $A_{f}$ are the probability amplitudes for the forward and backward creation of a photon pair; $\mathbf{p}_{i}$ is the pump annihilation operator in direction $i$; $\mathbf{fj}^{\dagger}$ and $\mathbf{bj}^{\dagger}$ are the forward and backward downconversion creation operators for events $j$, respectively; and $h.c.$ is the Hermitean conjugate. Taking the Taylor expansion, and retaining only terms where at least one pair of photons are emitted into both $f$ \& $b$ (emission into only one will not cause a trigger event and are disregarded), and assuming a large pump field so that we can ignore the annihilation of pump photons, we obtain the source terms,
\begin{eqnarray}
& A_{f}\, A_{b} \mathbf{b1}^{\dagger} \mathbf{b2}^{\dagger} \mathbf{f1}^{\dagger} \mathbf{f2}^{\dagger} + \notag \\
& (A_{f} A_{b}^2 \mathbf{b1}^{\dagger 2} \mathbf{b2}^{\dagger 2} \mathbf{f1}^{\dagger} \mathbf{f2}^{\dagger} + A_{f}^2 A_{b} \mathbf{b1}^{\dagger} \mathbf{b2}^{\dagger} \mathbf{f1}^{\dagger 2} \mathbf{f2}^{\dagger 2})/2.
\label{sourceeq}
\end{eqnarray}
The first term describes creation of one pair of photons into each the forward and backward modes, whereas the last two terms describe production of 2+1 and 1+2 photon pairs respectively. The contribution from terms higher than these have been found to be negligible and hence are ignored.

We encode logical $0$ \& $1$ in vertical and horizontal polarisations, $V$ \& $H$, respectively. The input modes to the gate, $\mathbf{f1}^{\dagger}$ and $\mathbf{b1}^{\dagger}$, are projected with a polarising beamsplitter, $\mathbf{f1}^{\dagger} {\rightarrow}  \alpha \, \mathbf{a}_H^{\dagger} {+} \beta \,\mathbf{a}_V^{\dagger}$, $\mathbf{b1}^{\dagger} {\rightarrow} \sigma \,\mathbf{b}_H^{\dagger} {+} \tau \,\mathbf{b}_V^{\dagger}$,
and the action of the controlled-\textsc{z} gate is described by,
\begin{eqnarray}
\mathbf{a}_{i}^{\dagger} & \rightarrow & -\sqrt{\eta_{i}}\, \mathbf{c}_{i}^{\dagger}{+} \sqrt{1-\eta_{i}}\, \mathbf{d}_{i}^{\dagger} ,\notag \\
\mathbf{b}_{i}^{\dagger} & \rightarrow & \sqrt{\eta_{i}}\, \mathbf{d}_{i}^{\dagger} {+} \sqrt{1-\eta_{i}}\, \mathbf{c}_{i}^{\dagger},
\end{eqnarray}
where $i$=$H,V$. Ideally, the reflection probabilities are $\eta_{H}$=$1/3$ and $\eta_{V}$=$1$, implementing the maximally entangling controlled-sign operation. Photon loss is modelled by additional beamsplitters, with reflectivity $k_{i}$. 

Measurement of both the trigger and gate photons is modelled by ideal projective measurements with non-number-resolving detectors. A quantitative comparison of our model and experiment requires values for the downconversion amplitudes, $A_{f}$, $A_{b}$, the beamsplitter reflectivities, $\eta_{i}$, and the losses, $k_{i}$. The reflectivities are easily obtained from direct measurement either with a suitable laser or the down-converted photons themselves. The other values need to be derived from specific measurements made using input states designed to extract these quantities.

\begin{table}[!t]
\begin{tabular} {|lc|c|c|}
\hline
Quantity & Variable & Ideal & Exp \\
\hline
\hline
PPBS reflectivity for H & $\eta_{H}$ & $1/3$ & $0.35$\\
PPBS reflectivity for V & $\eta_{V}$ & $1$ & $0.99$\\
Loss probability, mode f2 & $k_{f2}$ & $0$ & $0.904$\\
Loss probability, mode b2 & $k_{b2}$ & $0$ & $0.911$\\
Loss probability, mode c & $k_{c}$ & $0 $& $0.953$\\
Loss probability, mode d & $k_{d}$ & $0$ & $0.970$\\
Measured forward amplitude & $A_{f}$ & $0.137$ & $0.116$\\
Measured backward amplitude & $A_{b}$ & $0.208$ & $0.177$\\ \hline
\end{tabular}
\caption{Summary of values used as model inputs. When modelling the presence of a given error the experimental value is utilised; the ideal value is used to switch the error off. Uncertainties are in the last significant figure. \vspace{-4mm}
}
\label{values}
\end{table}

A summary of all values is given in Table \ref{values}. To derive the down-conversion amplitude values, consider the situation where one pair of photons is emitted in both the forward and backward directions. The probability of detecting a fourfold event with measurement settings $\{r,s\}$, where $r, s \in \{H,V,D,A,R,L\}$, is,
\begin{equation}
\vec{P}^{11} = A_{f}^2 A_{b}^2 \Bigg( \prod_{i=1}^{4} (1-k_{i}) \Bigg) \vec{\gamma}^{11},
\label{P11}
\end{equation}
where the superscript on $P$ refers to the photon number in each gate input, $a$ \& $b$;  $k_{i}$ is the probability of photon loss in modes $\{i\}$=$\{f2,c,d,b2\}$, and $\vec{\gamma}^{11}$ is the vector of overlap probabilities between the gate output and the measurement setting $rs$, $\vec{\gamma}^{11}$=$\{\langle rs | \mathbf{U}^{\dagger}_{gate} | \psi_{in} \rangle \}$. Without loss of generality we choose $| \psi_{in} \rangle$=$\mathbf{a}^{\dagger}_{D} \mathbf{b}^{\dagger}_{D}|00\rangle$ in the following discussion, so as to equally populate the logical states.

Experimentally it is tempting to obtain $\vec{P}^{11}$ by inputting $| DD\rangle$, and forming a vector of the resulting probabilities, $P^{11}$=$\{C_{rs}/C_{tot}\}$, where $C$ are counts and $C_{tot}$ is the number of total counts in the appropriate complete set of projectors (i.e. $C_{HH}$+$C_{HV}$+$C_{VH}$+$C_{VV}$, $C_{DD}$+$C_{DA}$+$C_{AD}$+$C_{AA}$ ...). However this does not account for events where 2 pairs of photons are emitted in the one direction and 1 pair in the other---a non-negligible background. These terms cannot be measured directly. However they can be estimated in the following manner. Consider the 2 pair emission in one direction, e.g. forward, and stop the photons from the other direction from entering the gate, e.g block mode $b$. The probability of detecting a fourfold event now is,
\begin{eqnarray}
\vec{P}^{20} & = & 
\frac{1}{4} A_{f}^4 A_{b}^2 \Bigg( \prod_{i=1}^{4} (1-k_{i}) \Bigg) (1+k_{f2}) \, \vec{\gamma}^{20},
\label{P20}
\end{eqnarray}
where $\vec{\gamma}^{20}$=$\{\langle rs | \mathbf{U}^{\dagger}_{gate} | \psi'_{in} \rangle \}$ and $| \psi'_{in} \rangle$=$\mathbf{a}^{\dagger}_{D} \mathbf{a}^{\dagger}_{D} \mathbf{b}^{\dagger}_{D}|00\rangle$. (Swapping the roles of the forward and backward directions gives $\vec{P}^{02}$ \& $\vec{\gamma}^{02}$). Experimentally we obtain $\vec{P}^{20}$ \& $\vec{P}^{02}$ by blocking in turn one of the gate inputs, while continuing to count four-fold events---since there is only one gate input active at a time, and both gate detectors fire, two photons must have been injected in the same input. In the case of perfect detection efficiency, the total number of events where two-forward and one-backward pairs are created is $N^{20}$=$\vec{C}^{20} \vec{\gamma}^{20}$. This of course is the same whether mode $b$ is blocked or not, i.e. $N^{20}$=$N^{21}$ and,
\begin{eqnarray}
\vec{C}^{20} \vec{\gamma}^{20} & = & \vec{C}^{21} \vec{\gamma}^{21}.
\label{N=N}
\end{eqnarray}
From the ratio of eqns \ref{P20} \& \ref{P11} the forward amplitude is,
\begin{eqnarray}
A_{f}^{2} & = & \frac{1}{1+k_{f2}} \sum^{4}_{r=1} \sum^{4}_{s=1} \frac{P^{20}_{rs} \gamma^{11}_{rs} }{ P^{11}_{rs} \gamma^{20}_{rs} }.
\end{eqnarray}
Remembering that $P_{rs}$=$C_{rs}/C_{tot}$, this becomes,
\begin{widetext}
\begin{eqnarray}
A_{f}^{2} & = & \frac{1}{1+k_{f2}} \sum^{4}_{r=1} \sum^{4}_{s=1} \frac{C^{20}_{rs} \gamma^{11}_{rs} }{ C^{11}_{rs} \gamma^{20}_{rs} } = \frac{1}{1+k_{f2}} \sum^{4}_{r=1} \sum^{4}_{s=1} \frac{C^{20}_{rs} \gamma^{11}_{rs} }{ (C'^{11}_{rs}-C^{21}_{rs}-C^{12}_{rs}) \gamma^{20}_{rs} },
\end{eqnarray}
\end{widetext}
where $C'^{11}$ is the measured number of four-fold events, $C'^{11}$=$C^{11}$+$C^{21}$+$C^{12}$ and, from equality \ref{N=N}, $C^{21}_{rs}$=$C^{20}_{rs} \gamma^{20}_{rs}/\gamma^{21}_{rs}$ and similarly for $C^{12}$. (Swapping the forward and backward roles in the above argument yields $A_{b}$). From our measurements we determined $A_{f}$=0.137 and $A_{b}$=0.208 in the no-loss limit, $k_{i}$=0. Note that in the high loss limit, $k_{i}$$\rightarrow$1, and absence of single photon sources or number resolving detectors our estimate of the downconversion amplitudes  $A_{i}$ will decrease by a factor of $\sqrt{2}$. This somewhat counter-intuitive result highlights the critical role of loss in the presence of unsuppressed higher-order photon terms: the combination causes errors, in this case an overestimation of the downconversion probability.

We estimated the loss probabilities of our experiment using the following method. We input a pair of vertically-polarised photons, which ideally both reflect from the PPBS, and measure with the VV setting. We measured the singles rate of each detector, $S_{i}$ and the two-fold coincidences, $C_{12}$ \& $C_{34}$, caused respectively by pairs generated in the forward and backward directions. Accounting for background singles counts, $B$, and coincidence accidental counts, $A$, the loss in mode $i$ is,
\begin{equation}
k_{i} = 1- \frac{C_{ij}-A_{ij}}{S_{j}-B_{j}}
\end{equation}
where $i,j$ $\in$ $\{1,2\}$ or $\{3,4\}$. We are clearly in a high-loss regime, $k_{f2}$=$0.904$, $k_{c}$=$0.953$, $k_{d}$=$0.970$, and $k_{b2}$=$0.911$---the downconversion amplitudes are $A_{f}$=0.116 and $A_{b}$=0.177.